\documentclass[aps,prb,english,twocolumn,showpacs,preprintnumbers,amsmath,amssymb,floatfix,superscriptaddress,longbibliography]{revtex4-1}

\usepackage[T1]{fontenc}
\usepackage[latin9]{inputenc}
\usepackage{graphicx}
\usepackage{amssymb}
\usepackage{babel}
\makeatletter

\usepackage[version=3]{mhchem}

\usepackage{color}

\global\arraycolsep=2pt
\usepackage{stmaryrd}
\usepackage{amsmath}
\usepackage{amssymb}
\usepackage{graphicx}
\usepackage{textcomp}
\usepackage{calrsfs}
\usepackage{yfonts}
\usepackage{bm}
\usepackage{color}
\newcommand{\ben}{\begin{equation*}}
\newcommand{\een}{\end{equation*}}
\newcommand{\bean}{\begin{eqnarray*}}
\newcommand{\eean}{\end{eqnarray*}}

\newcommand{\be}{\begin{equation}}
\newcommand{\ee}{\end{equation}}
\newcommand{\bea}{\begin{eqnarray}}
\newcommand{\eea}{\end{eqnarray}}

\newcommand{\psumbar}{\sum\nolimits^\prime}
\DeclareMathOperator*{\psum}{\psumbar}

\makeatother
\usepackage{babel}
\makeatother

\begin{document}
\title{Premelting and formation of ice due to Casimir-Lifshitz interactions: Impact of improved parameterization for materials}

\author{Yang Li}
  \email{leon@ncu.edu.cn}
  \affiliation{Department of Physics, Nanchang University, Nanchang 330031, China}
  \affiliation{Institute of Space Science and Technology, Nanchang University, Nanchang 330031, China}

 \author{Kimball A. Milton}
  \email{kmilton@ou.edu}
  \affiliation{Homer L. Dodge Department of Physics and Astronomy, University of Oklahoma, Norman, Oklahoma 73019, USA}

 \author{Iver Brevik}
  \email{iver.h.brevik@gmail.com}
  \affiliation{Department of Energy and Process Engineering, Norwegian University of Science and Technology, NO-7491 Trondheim, Norway}

\author{Oleksandr I. Malyi}
  \affiliation{Renewable and Sustainable Energy Institute, University of Colorado 4001 Discovery Drive, Boulder, CO 80309-029, USA}


\author{Priyandarshini Thiyam}
  \affiliation{Stranski-Laboratorium f\"ur Physikalische und Theoretische Chemie, Institut f\"ur Chemie, Technische Universit\"at Berlin, 10623 Berlin, Germany}

\author{Clas Persson}
  \affiliation{Centre for Materials Science and Nanotechnology, Department of Physics, University of Oslo, P. O. Box 1048 Blindern, NO-0316 Oslo, Norway}
\affiliation{Department of Materials Science and Engineering, Royal Institute of Technology, SE-100 44 Stockholm, Sweden}

\author{Drew F. Parsons}
  \affiliation{Department of Chemical and Geological Sciences,University of Cagliari, Cittadella Universitaria, 09042 Monserrato, CA, Italy}

\author{Mathias Bostr{\"o}m}
 \email{mathias.bostrom@smn.uio.no}
  \affiliation{Centre for Materials Science and Nanotechnology, Department of Physics, University of Oslo, P. O. Box 1048 Blindern, NO-0316 Oslo, Norway}

\begin{abstract}
Recently, the premelting and formation of ice due to the Casimir-Lifshitz interaction, proposed in early 1990s by Elbaum and Schick [Phys. Rev. Lett. 66, 1713-1716 (1991)], have been generalized to diverse practical scenarios, yielding novel physical intuitions and possibilities of application for those phenomena. The properties of materials, in particular, the electrical permittivity and permeability, exert significant influences on the Casimir-Lifshitz energies and forces, and hence on the corresponding premelting and formation of ice. To address these influences in detail and explore the resulting physics, here we revisit and extend the analyses of previous work, with both the dielectric data utilized there and the latest dielectric functions for ice and cold water. While our previous results are rederived, an error there has also been spotted. For the four-layer cases considered by some of us, the existence of stable configurations depending on the initial conditions has been confirmed, and different types of stability corresponding to minima of the Casimir-Lifshitz free energy are demonstrated. As the new dielectric functions for ice and cold water deviate considerably from those used by Elbaum and Schick, their vital impacts on three- and four-layer configurations are therefore being reconsidered.
\end{abstract}

\date{\today}

\maketitle

\section{Introduction}
\label{I}
\par Relevant surface free energies should be carefully evaluated when considering the formation of ice and water layers. An important contribution to the theories for surface forces came from Casimir,\cite{casimir1948-1} who related the force between a pair of neutral planar perfect metal surfaces to changes in the zero-point energy of vacuum. Later, a much more general theory was derived by Lifshitz and co-workers for interactions between surfaces.\cite{Lifshitz1956,Dzya} Attempts to verify the Lifshitz formula\cite{Lifshitz1956,Dzya} were taken quite early using surface force measurements,\cite{TaborWinterton1969,IsraelachviliTabor1972,WhiteIsraelachviliNinham1976} although those were far from conclusive. Much more definitive were studies of thin film growth.\cite{AndSab,Haux,sabisky1973verification} It was in parallel extensively studied theoretically, for instance by the groups led by Parsegian and Ninham.\cite{ParsegianNinham1969,ninham1970van,richmond1971calculations,richmond1971note,Ninham_2019} Since the late 1990s, accurate measurements for Casimir forces between metal surfaces were carried out by Lamoreaux and co-workers,\cite{Lamo1997,Lamo1998,SushNP} and later by others.\cite{klimchitskayaMostepanenko_Casimirforce2020,LiuZhangKlimchitskayaCasimirGraphene2021,decca21}   The zero-temperature theory has been rather well verified, but the finite-temperature correction, especially for systems consisting of imperfect metals, remains controversial.\cite{bostrom2000thermal,breviknjp,brevik2013temperature,klimchitskaya2017casimir,sernelius2018fundamentals,klimchitskayaMostepanenko_Casimirforce2020,LiuZhangKlimchitskayaCasimirGraphene2021,decca21} The books by Milton,\cite{milton2001casimir} Bordag et al.,\cite{Bordagbook} Dalvit et al.,\cite{dalvit2011casimir} Buhmann,\cite{buhmann2013dispersion1,buhmann2013dispersion2} and Sernelius,\cite{sernelius2018fundamentals} describe well this active and
important research topic.

\par Notably, interesting phenomena and applications also come from the finite-temperature Casimir-Lifshitz formula. For instance, the potential role of Casimir-Lifshitz free energy in ice premelting\cite{elbaum1991app} and formation\cite{Elbaum2} was put forward by Elbaum and Schick. In the specific context of ice/water systems, it has in particular been shown that Casimir-Lifshitz forces,\cite{elbaum1991app,Elbaum2,Wilen,Thiyam2016,Bostr2016,MBPhysRevB02017,Prachi2019role}  combined with double layer forces due to experimentally unknown impurity charges,\cite{Wilen,Wettlaufer,ThiyamFiedlerBuhmannPerssonBrevikBostromParsons2018} can give rise to short range repulsion and long range attraction leading to an equilibrium system with the experimentally observable partial ice premelting.\cite{Dash_surfacemelting,Elbaum1993,Dash1995,RevModPhys.78.695,li2019interfacial} The existence of thin surface films of liquid water on ice particles\cite{benet16,benet19} has been proposed as a factor influencing environmental physics, including potential effects on frost heave,\cite{WilenFrost} and charging of thunderclouds.\cite{Baker,DW,SPWettlaufer2006} Based on experimental observations, heat insulating gas hydrate strata was proposed to play a role for the internal geophysics on Pluto\cite{KamataNimmoSekineKuramotoNoguchiKimuraTani2019} and the moon Enceladus.\cite{MunozPrietoBallesteros2021} Both the chemical composition and insulating properties for large scale heat insulating strata on, e.g., the moon Enceladus could be influenced by formation of Casimir-Lifshitz energy induced micronsized ice layers on the interface between gas hydrate clusters and the salty ocean water.\cite{Bostrom2019dispersion,BostrometalAA2021}

\par In an expansion of the work by Elbaum and Schick,\cite{elbaum1991app,Elbaum2} we predicted that ice could form at a water-silica interface at the triple point of water.\cite{MBPhysRevB02017} This is an update and expansion of that paper,
and the purpose is four-fold. Firstly, we will correct a minor error of ours in Ref.~\onlinecite{MBPhysRevB02017}. This brings about a better agreement between results obtained by using \emph{ab initio} calculated dielectric functions for silica materials~\cite{Sasha2016} and those using experimental dielectric functions (the latter derived from a paper by van Zwol and Palasantzas,\cite{Zwol2010} but treated incorrectly in Ref.~\onlinecite{MBPhysRevB02017}). Secondly, we will compare our past predictions about the ice formation on silica surfaces, which are based on ice and water dielectric functions from Elbaum and Schick,\cite{elbaum1991app} with predictions based on different revised and improved dielectric functions for ice~\cite{luengo2021lifshitz,luengo2021WaterIce} and water~\cite{JohannesWater2019,luengo2021lifshitz,luengo2021WaterIce} at the triple point of water. Thirdly, this also encourages us to take a careful look at the results obtained by Elbaum and Schick for ice premelting~\cite{elbaum1991app} and ice formation on water surfaces.\cite{Elbaum2} The contradictory results found deserve careful inspection, and the significant dependence on the dielectric functions of media should be emphasized. Finally, we use a new theory for Casimir-Lifshitz free energies in inhomogeneous media~\cite{LiMiltonGuoKennedyFulling2019} to explore some generalized cases beyond the three-layer system described by the classical Lifshitz theory.

\par This work is organized as follows: In Sec.~\ref{T}, we present the theory, including the generalized case where more than one layer are allowed to expand at the expense of other layers,\cite{esteso2020premelting,luengo2021lifshitz} and we also discuss relevant configurations. As has been demonstrated in our past work,\cite{MBPhysRevB02017} with the \emph{ab initio} dielectric functions for silica materials\cite{Sasha2016} and the experimentally based dielectric functions from Elbaum and Schick\cite{elbaum1991app,Elbaum2} for ice and water, an ice layer could grow between the water and silica interface. Within the models applied, this previous result is correct as shown in Sec.~\ref{R}, but additional results using two different data sets for silica based on parameterizations from van Zwol and Palasantzas\cite{Zwol2010} were in error due to an incorrect numerical treatment. We will demonstrate that when this is corrected, plausible results follow, which are consistent with the results\cite{MBPhysRevB02017} obtained using silica dielectric functions from both Malyi et al.\cite{Sasha2016} and Grabbe.\cite{Grabbe1993} We first correct, then explore and expand both our past work~\cite{MBPhysRevB02017} and those of Elbaum and Schick\cite{elbaum1991app,Elbaum2} in the light of the new improved dielectric functions for ice and water. In Sec.~\ref{D}, we try to provide some detailed discussions on the results demonstrated in Sec.~\ref{R}. A simple characteristic parameter method is proposed to facilitate the estimation of the stability properties influenced by the Casimir-Lifshitz interaction. Conclusions are given in Sec.~\ref{C}.


\begin{figure*}[htp]
  \centering
  \includegraphics[scale=0.35]{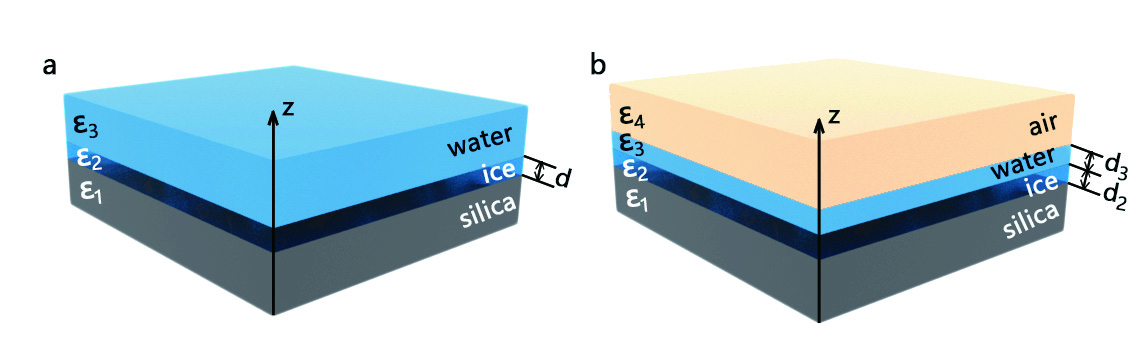}
  \caption{\label{fig.finalsiice_0906}Schematic illustrations for three- and four-layer configurations. (\textbf{a}) A example of three-layer configuration, namely silica-ice-water. (\textbf{b}) A example of four-layer configuration, namely silica-ice-water-vapor.}
\end{figure*}

\section{Theory}
\label{T}
\par The natural units $\hbar=c=\varepsilon_0=\mu_0=k_B=1$ are used, unless specified in the current publication. In our previous work,\cite{MBPhysRevB02017} we considered a planar layered system consisting of silica, ice and water. The Casimir-Lifshitz free energy in such a three-layered planar geometry (Dzyaloshinskii-Lifshitz-Pitaevskii or DLP configuration for short), where medium 1 and 3 are separated by medium 2 with a thickness $d$ (for example Fig.~\ref{fig.finalsiice_0906}a), is well known.\cite{Lifshitz1956,Dzya}
\begin{eqnarray}
\label{eqT.1}
F
=
T\psum_{n=0}^{\infty}\int\frac{d^2k}{(2\pi)^2}
\sum_{s=\rm E,H}
\ln(1+r^{s}_{32}r^{s}_{21}e^{-2\kappa_2d})
.
\end{eqnarray}
The reflection coefficients for transverse electric (TE, $s=\rm E$) are defined as
\begin{eqnarray}
\label{eqT.2}
r^{\rm E}_{ij}=\frac{
\kappa_i\mu_j-\kappa_j\mu_i
}{
\kappa_i\mu_j+\kappa_j\mu_i
},\
\kappa_i=\sqrt{k^2+\varepsilon_i(i\zeta_n)\mu_i(i\zeta_n)\zeta_n^2},
\end{eqnarray}
in which the electromagnetic response properties of medium $i$ is described by the permittivity $\varepsilon_i$ and permeability $\mu_i$. Here, $\zeta_n=2\pi Tn$ is the Matsubara frequency. The transverse magnetic (TM, $s=\rm H$) counterpart of $r^{\rm E}_{ij}$ is obtained by making the substitution $\varepsilon\leftrightarrow\mu$. In this paper, we focus on the nonmagnetic media.

\par Very recently, Parashar et al.,\cite{ParasharQEstress2018} Li et al.,\cite{LiMiltonGuoKennedyFulling2019} Esteso et al.,\cite{esteso2020premelting} Luengo-M{\'a}rquez and MacDowell,\cite{luengo2021lifshitz,luengo2021WaterIce} demonstrated how to expand the classical Lifshitz theory to the case containing two intermediate layers that can both grow or decrease at the expense of each other. For the four-layer system illustrated by Fig.~\ref{fig.finalsiice_0906}b, the Casimir-Lifshitz free energy between medium 1 and 4 is derived as (for an alternative
derivation, please refer to the Appendix)
\begin{eqnarray}
\label{eqT.3}
F_{14}
&=&
T\psum_{n=0}^{\infty}\int\frac{d^2k}{(2\pi)^2}
\sum_{s=\rm E,H}
\ln\bigg[
1+\frac{(1+r_{32}^s)r_{43}^se^{-2\kappa_3d_3}}{1+r_{43}^sr_{32}^se^{-2\kappa_3d_3}}
\nonumber\\
& &
\times
\frac{(1-r_{32}^s)r_{21}^se^{-2\kappa_2d_2}}{1+r_{32}^sr_{21}^se^{-2\kappa_2d_2}}
\bigg]
,
\end{eqnarray}
where $d_2$ and $d_3$ are the thicknesses of media 2 and 3, respectively. When media 2 and 3 are the same, then the reflection coefficients $r_{32}^{\rm E}$ and $r_{32}^{\rm H}$ are both zero, which means Eq.~\eqref{eqT.3} reduces to the classical DLP free energy as in Eq.~\eqref{eqT.2}. When Casimir-Lifshitz free energies of the the slab 1-2-3 and 2-3-4 DLP configurations are included, then the total free energy $F=F_{14}+F_{13}+F_{24}$ is expressed as
\begin{eqnarray}
\label{eqT.4}
F
&=&
T\psum_{n=0}^{\infty}\int\frac{d^2k}{(2\pi)^2}
\sum_{s=\rm E,H}
\ln(
1+r_{43}^sr_{32}^se^{-2\kappa_3d_3}
\nonumber\\
& &
+r_{32}^sr_{21}^se^{-2\kappa_2d_2}
+r_{43}^sr_{21}^se^{-2\kappa_3d_3-2\kappa_2d_2}
),
\end{eqnarray}
which is consistent with the results presented by Esteso et al.\cite{esteso2020premelting} and more recently by Luengo-M{\'a}rquez and MacDowell.\cite{luengo2021lifshitz} (Since in the 4-media case there are three reflection coefficients, one might think that terms with three reflection coefficients could appear. Such terms, however, are forbidden by parity invariance; only even powers of reflection coefficients can appear in the free energy.)

\par According to Eq.~\eqref{eqT.1} and Eq.~\eqref{eqT.2}, when conditions, such as $r_{32},r_{21}>0$, are properly satisfied so that the Casimir-Lifshitz free energy depends on the separation between medium 1 and 3 as a monotonically decreasing function, then a repulsive Casimir-Lifshitz force is found. This implies that the expansion of the
intermediate medium 2 decreases the free energy, forming a more stable system. As has been known since Ref.~\onlinecite{Dzya}, this kind of condition could typically be satisfied if $\varepsilon_1>\varepsilon_2>\varepsilon_3$ or $\varepsilon_1<\varepsilon_2<\varepsilon_3$. This repulsion has already been observed in
experiments.\cite{sabisky1973verification,Munday2009}
Similarly, if the free energy increases with the separation, an attractive Casimir-Lifshitz force results, which means the system would tend to eliminate the intervening medium to stabilize itself. Therefore, the electromagnetic response properties of media will be significant to the stability induced by the fluctuating electromagnetic field. Moreover, when two bulk media interact via more than one intervening medium, multiple stabilities could show up, implying more complicated possibilities and phenomena.

\par In the following section, some realistic cases will be investigated to acquire an explicit understanding of the role of Casimir-Lifshitz free energy in the premelting or formation of ice in these cases and its generalizations.

\section{Results}
\label{R}
\subsection{Results using Elbaum and Schick models for ice and water}
\label{RES}
\par
\begin{figure}[htp]
  \centering
  \includegraphics[scale=0.35]{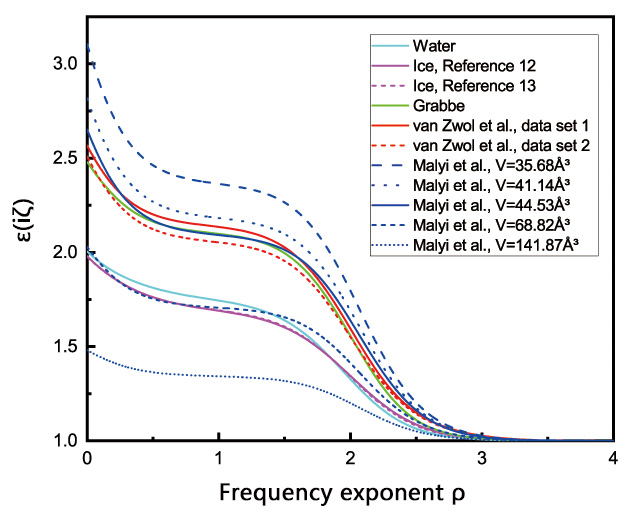}
  \caption{\label{fig.A10}Permittivities of water (Cyan), ice (Magenta) and different silica materials (green, red and blue for Grabbe,\cite{Grabbe1993} van Zwol \emph{et al.}\cite{Zwol2010} and Malyi \emph{et al.},\cite{Sasha2016} respectively) as functions of frequency exponent $\rho$. The frequency (in units of $\rm rad/s$) $\zeta$ is related to the exponent $\rho=\log_{10}\zeta/\zeta_T$, in which $\zeta_T=2\pi T$ is the $n=1$ Matsubara frequency at the temperature of the triple-point of water, $273.16\,\rm K$. The static dielectric constants for ice and water are, respectively, 91.5 and 88.2 as in Ref.~\onlinecite{elbaum1991app}, while for $\rm SiO_2$ materials, their static values are $4.97$, $4.32$, $3.90$, $2.62$, and $1.69$ from Ref.~\onlinecite{Sasha2016} for average volumes per $\rm SiO_2$ $V=35.68$, $41.14$, $44.53$, $68.82$, and $141.87$\,{\AA}$^3$, respectively, $3.80$ for Grabbe,\cite{Grabbe1993} and $3.90$ for data set 1 and data set 2 of van Zwol \emph{et al.}.\cite{Zwol2010} The UV contributions of the ice dielectric function are modeled according to Refs.~\onlinecite{daniels1971bestimmung,seki1981optical}, which are, respectively, Reference 12 and Reference 13 quoted in Ref.~\onlinecite{elbaum1991app}.}
\end{figure}

 In this part, we recheck and extend some previous work. In Ref.~\onlinecite{MBPhysRevB02017}, we studied the ice growth on the interface of silica and water, i.e. the silica-ice-water system, at the triple-point of water. The dielectric functions for ice and water are taken from Elbaum and Schick.\cite{elbaum1991app,Elbaum2} As for dielectric functions of silica with various nanoporosities, previous first-principle evaluation,\cite{Sasha2016} together with the phonon contributions counted, is employed.\cite{MBPhysRevB02017} To make comparisons, we also used silica data from Grabbe\cite{Grabbe1993} and from van Zwol and Palasantzas.\cite{Zwol2010} The dispersions of dielectric functions of ice, water and silica materials are plotted in Fig.~\ref{fig.A10}, where the error in the calculation when using the data from van Zwol and Palasantzas\cite{Zwol2010} in our previous paper\cite{MBPhysRevB02017} has been corrected.

\par
\begin{figure}[htp]
  \centering
  \includegraphics[scale=0.32]{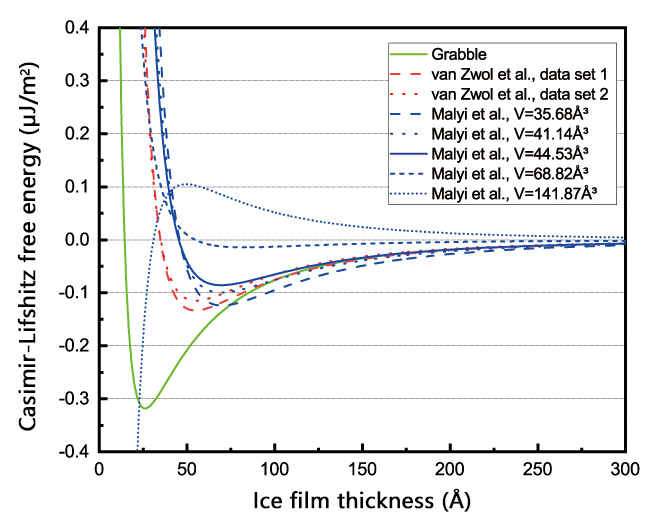}
  \caption{\label{fig.A12}The Casimir-Lifshitz free energy per unit area for silica-ice-water configuration at the triple-point of water as a function of the thickness of ice layer. Its dependence on silica permittivities, thus the nanoporosities, is shown.}
\end{figure}

\begin{figure}[htp]
  \centering
  \includegraphics[scale=0.33]{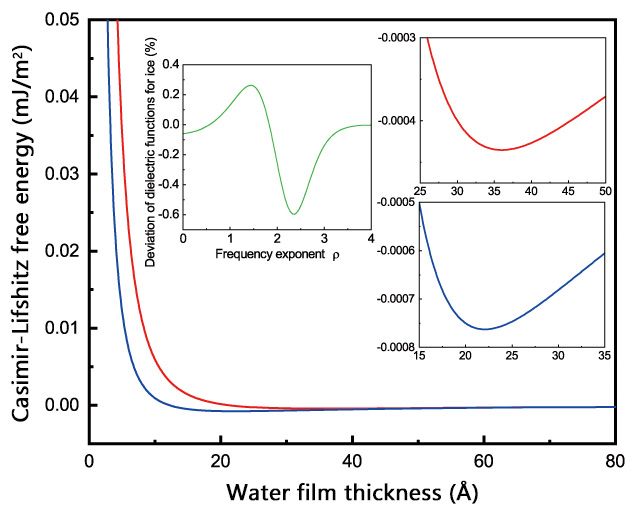}
  \caption{\label{fig.A1113}The Casimir-Lifshitz free energy of ice-water-vapor system as a function of water layer thickness. The results are obtained based on the data used by Elbaum and Schick in Ref.~\onlinecite{elbaum1991app}. The Casimir-Lifshitz free energies are plotted with the UV models for the ice dielectric function given by Reference 12 (red) and Reference 13 (blue) cited in Ref.~\onlinecite{elbaum1991app}, respectively.
  The behavior near the minima is shown in the right insets.
  The deviation between the dielectric functions for ice with different UV models, denoted $\varepsilon_{12}$ and $\varepsilon_{13}$, is defined as $(\varepsilon_{13}-\varepsilon_{12})/\varepsilon_{12}$. Its dependence on the frequency exponent $\rho$ is given in the middle inset.}
\end{figure}

 With the corrected data, Casimir-Lifshitz free energies of silica-ice-water systems with various silica materials are demonstrated in Fig.~\ref{fig.A12}. Although the corrections do not qualitatively change our result, that is, a nano-sized ice film could form between water and silica, quantitatively undeniable variations happen, for instance the magnitude of the minimum free energy for the data set 1 of van Zwol et al.\ is only about one third of that in previous work.\cite{MBPhysRevB02017} Actually this quantitative sensitivity with the properties of  material (mainly the permittivity in our study here) is not uncommon and deserves considerable care. The ice-water-vapor system investigated by Elbaum and Schick\cite{elbaum1991app,Elbaum2} is a good example.
As shown in Fig.~\ref{fig.A1113}, the results by Elbaum and Schick~\cite{elbaum1991app} have been rederived with the ice modeled according to Reference 12 cited in Ref.~\onlinecite{elbaum1991app} or Ref.~\onlinecite{daniels1971bestimmung}. The minimum of the Casimir-Lifshitz free energy is reached when the thickness of water layer is about $36$\,{\AA}. However, when the
Reference 13 or Ref.~\onlinecite{seki1981optical} model of ice in the ultraviolet region is employed, the premelting water layer still appears, but with a thickness around $22$\,{\AA}. Given the fact that the deviation between the two dielectric constant models for ice is less than $1\%$, the deviations resulting from differences of these two UV models for ice are quite striking. It is thus clear that to achieve an accurate quantitative predictions for the experimental exploration, highly reliable data about the dielectric functions of materials involved are crucial.

\par
\begin{figure*}[htp]
  \centering
  \includegraphics[scale=0.24]{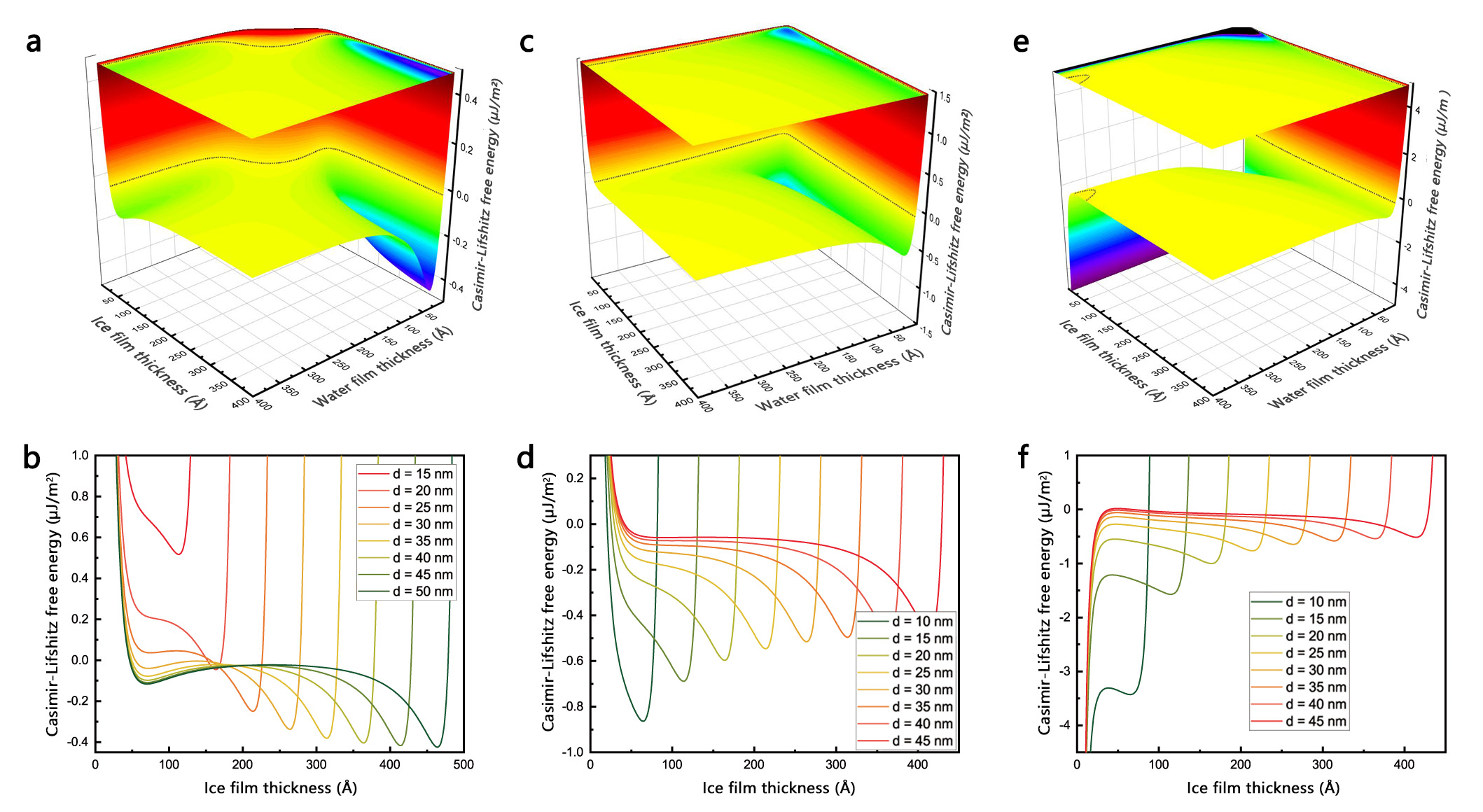}
  \caption{\label{fig.FigES4L}The Casimir-Lifshitz interaction free energy of silica-ice-water-vapor system at $T=273.16\rm K$ with the average volume per $\rm SiO_2$ being $V=35.68$\,{\AA}$^3$ (a and b), $V=68.82$\,{\AA}$^3$ (c and d) and $V=141.87$\,{\AA}$^3$ (e and f). For subfigures a, c and e (the corresponding projection of each 3D surface is given at the top), the dependence of free energy on the thicknesses of ice and water films are demonstrated. For subfigures b, d and f, the dependence of free energy as the function of ice film thickness, with the total thickness of ice and water layers fixed as $d$, are given.}
\end{figure*}

 Apart from the electromagnetic properties, the structure of the system can also provide remarkable diversities.  Esteso et al.\cite{esteso2020premelting} have investigated the premelting of a thin ice layer absorbed on a quartz rock surface through the stress tensor approach. The consistency with Refs.~\onlinecite{elbaum1991app,MBPhysRevB02017} justified their method, and multiple stable states for a given total thickness of ice and water layers were seen. Here we utilize the free energy approach in the hope to clarify relevant properties of the silica-ice-water-vapor system. As shown in Fig.~\ref{fig.FigES4L}a, for the silica material with the average cell volume $V=35.68$\,{\AA}$^3$, in the $d_2$-$d_3$ plane (here $d_2$ and $d_3$ are the thicknesses of ice and water layers, respectively), there are two regions, i.e., the green and blue trenches, where the Casimir-Lifshitz free energy minimizes itself locally. On other parts of that plane, however, the free energy will finally fall into either trench whenever possible, depending on the initial state. If the amount of water absorbed on this silica material is almost fixed, such that the total thickness of ice and water layers can be regarded as a constant, then the minimum of Casimir-Lifshitz free energy, as demonstrated in Fig.~\ref{fig.FigES4L}b, would be split into two by increasing total thickness $d=d_2+d_3$. The depths of the two minima increase much slower than does the distance between them, which may signify a clear separation between the two phases and facilitate possible experimental investigations.
If the average cell volume of $\rm SiO_2$ is increased to $V=68.82$\,{\AA}$^3$, then only one physically nontrivial minimum trench survives, and there is a global minimum of free energy on the $d_2$-$d_3$ plane (at about $d_2=3.2\,\rm nm$ and $d_3=3.9\,\rm nm$). In addition, by attaching more water to this silica material, the ice layers will keep growing and the water layer always remains in the nano-scale. If the average cell volume is further increased to about $V=141.87$\,{\AA}$^3$, the property of stability will be radically changed, as illustrated in Fig.~\ref{fig.FigES4L}e. The strong stability is now at $d_2=0$, which means then the system is stabilized when the ice layer vanishes. If an ice layer is put on this silica, its thickness will determine whether it could melt easily. On the other hand, when a water layer is on this silica, the Casimir free energy prevents the formation of ice at the silica-water interface.


\subsection{Results using recent models for ice and water}
\label{RLD}
Recently, Fiedler et al.\cite{JohannesWater2019} re-analysed the available data for the optical spectra for water. A careful parameterisation for ice cold water was performed, which simultaneously fitted both the real and imaginary parts of the dielectric function at real frequencies. Preliminary calculations, using the new dielectric function for water,\cite{JohannesWater2019} together with the older dielectric function for ice,\cite{elbaum1991app} resulted in a prediction that micron-sized ice layers can form at water surfaces. This is in contrast to the work by Elbaum and Schick,\cite{elbaum1991app,Elbaum2} in which the ice premelting~\cite{elbaum1991app} due to dispersion forces was predicted, while no ice formation on water surfaces\cite{Elbaum2} was seen. Considering the importance of this topic, we revisit this problem using the most recently reported dielectric functions for both ice and water obtained by Luengo-M{\'a}rquez and MacDowell.\cite{luengo2021lifshitz,luengo2021WaterIce}

\par
\begin{figure*}
  \centering
  \includegraphics[scale=0.32]{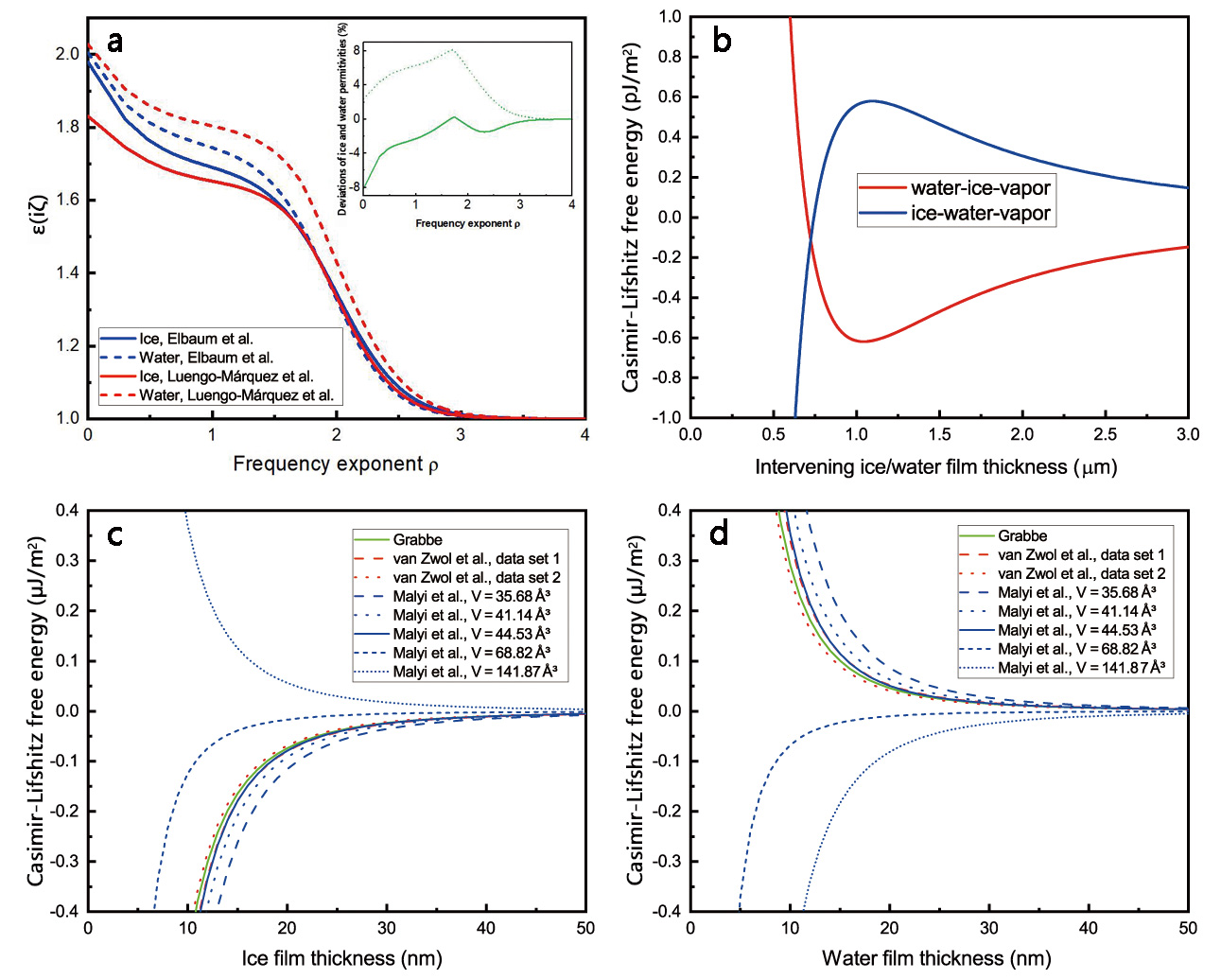}
  \caption{\label{fig.Fig_EpESLD}Previous work revisited with the latest dielectric functions for ice and water\cite{luengo2021lifshitz,luengo2021WaterIce}  near the triple point of water. (\textbf{a}) A comparison between the latest dielectric functions~\cite{luengo2021lifshitz,luengo2021WaterIce} (LM) and those used by Elbaum and Schick\cite{elbaum1991app,Elbaum2} (ES). The error between them is defined as $(\varepsilon_{\rm LM}-\varepsilon_{\rm ES})/\varepsilon_{\rm LM}$. The dependence of errors on the frequency exponent $\rho$ for ice (solid green) and water (dashed green) are given in the inset. (\textbf{b}) The Casimir-Lifshitz interaction free energies of ice-water-vapor and water-ice-vapor configurations evaluated with the latest data. Utilizing the new data, (\textbf{c}) and (\textbf{d}) demonstrate the dependences of free energy on ice and water films in the silica-ice-water and silica-water-ice configurations, respectively.}
\end{figure*}

\begin{figure*}[htp]
  \centering
  \includegraphics[scale=0.24]{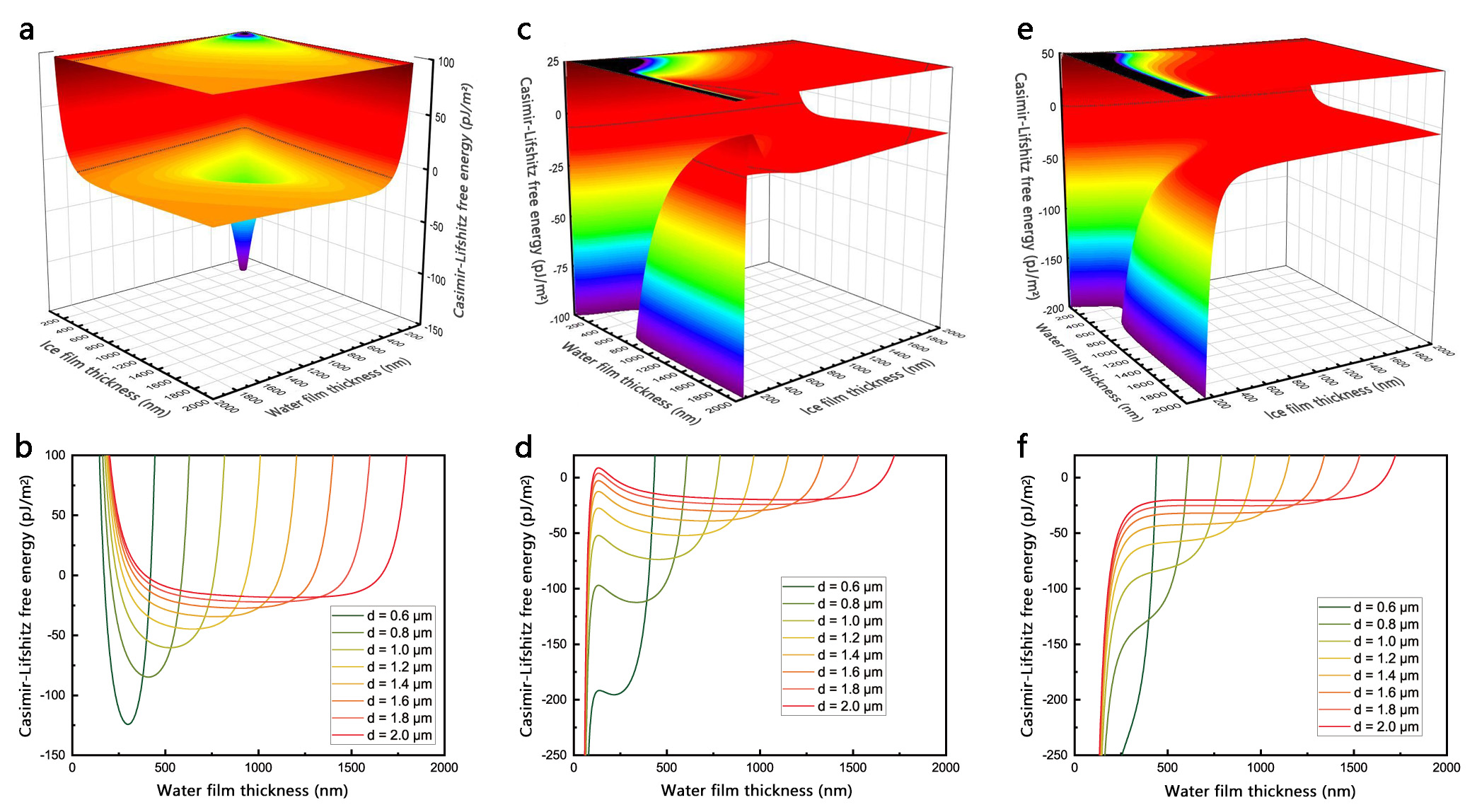}
  \caption{\label{fig.FigLD4L}The Casimir interaction free energy of silica-water-ice-vapor system at $T=273.16\,\rm K$ with the average volume per $\rm SiO_2$ being $V=35.68$\,{\AA}$^3$ (a and b), $V=68.82$\,{\AA}$^3$ (c and d) and $V=141.87$\,{\AA}$^3$ (e and f). The latest dielectric functions for ice and water from Luengo-M{\'a}rquez and MacDowell\cite{luengo2021lifshitz,luengo2021WaterIce} are employed. For subfigures a, c and e (the corresponding projection of each 3D surface is given at the top), the dependence of free energy on the thicknesses of ice and water films are demonstrated. For subfigures b, d and f, the dependence of free energy as the function of water film thickness, with the total thickness of ice and water layers fixed as $d=d_2+d_3$, are given.}
\end{figure*}

 As shown in Fig.~\ref{fig.Fig_EpESLD}a, the new dielectric functions deviate significantly from those used by Elbaum and Schick\cite{elbaum1991app,Elbaum2} (up to about $10\%$). To address the inconsistency between Fiedler et al.\cite{JohannesWater2019} and Elbaum and Schick,\cite{elbaum1991app} we evaluate Casimir-Lifshitz free energies in the ice-water-vapor and water-ice-vapor systems with these new data (see Fig.~\ref{fig.Fig_EpESLD}b). The results are clearly in favor of the conclusions in Ref.~\onlinecite{JohannesWater2019}, since an ice film of micron size is promoted by the Casimir-Lifshitz interaction, while the ice-water-vapor configuration is unstable. Therefore, it seems that the investigations are converging to similar conclusions as the properties of media are characterized  more and more accurately. Further exploring the ice-freezing on silica rocks as in Ref.~\onlinecite{MBPhysRevB02017}, we see a vastly different picture. As illustrated in Fig.~\ref{fig.Fig_EpESLD}c and Fig.~\ref{fig.Fig_EpESLD}d, with the dielectric functions for silica by Malyi et al.,\cite{Sasha2016} if the average cell volume of silica is small enough, for instance $V=35.68$\,{\AA}$^3$, the ice film separating this silica and water should vanish to stabilize the system. On the contrary, when the average cell volume $V$ is large, such as $V=141.87$\,{\AA}$^3$, a nano-sized water film is predicted to vanish. The sensitivity to the average cell volume is particularly striking. For example, for $V=68.82$\,{\AA}$^3$, as shown in Fig.~\ref{fig.Fig_EpESLD}, the material contacting with the surface of this silica can be either ice or water, depending on the initial state. In these three-layer cases, the premelting and formation of ice, characterized by the nano- or micron-sized intervening layer, are not seen with the new data, unlike that claimed before.\cite{MBPhysRevB02017} But the multi-stable property for some particular case may offer convenience for observations.
As for the four-layer cases, drastic modifications caused by the new data are obvious according to Fig.~\ref{fig.FigLD4L}. We see one and only one local and global minimum of Casimir-Lifshitz free energy (located at about $d_3\approx d_2=0.2\mu\rm m$) in Fig.~\ref{fig.FigLD4L}a, instead of two local minima for separated regions as in Fig.~\ref{fig.FigES4L}a. This is further illustrated by comparing Fig.~\ref{fig.FigLD4L}b with Fig.~\ref{fig.FigES4L}b. Furthermore, for the $V=68.82${\AA}$^3$ silica, Fig.~\ref{fig.FigLD4L}c and Fig.~\ref{fig.FigES4L}c show totally different behaviors. Without a trench as in Fig.~\ref{fig.FigES4L}c, Fig.~\ref{fig.FigLD4L}c is more analogous to Fig.~\ref{fig.FigES4L}e, and the similarities are also found between Fig.~\ref{fig.FigLD4L}d and Fig.~\ref{fig.FigES4L}f. The difference between the $V=68.82$\,{\AA}$^3$ cases in Fig.~\ref{fig.FigES4L}c and Fig.~\ref{fig.FigLD4L}c is that a much more deeper basin is found in Fig.~\ref{fig.FigLD4L}c. For rock built from quartz (the silica that mimics this best in our study is the one with $V=35.68$\,{\AA}$^3$), we find in Fig.~\ref{fig.FigLD4L}b a shallow energy minimum as a function of water layer thickness. This minimum moves with increasing total thickness of the ice+water layer indicating growth of the water layer.

\section{Discussion}
\label{D}
\par
\begin{figure*}[htp]
  \centering
  \includegraphics[scale=0.32]{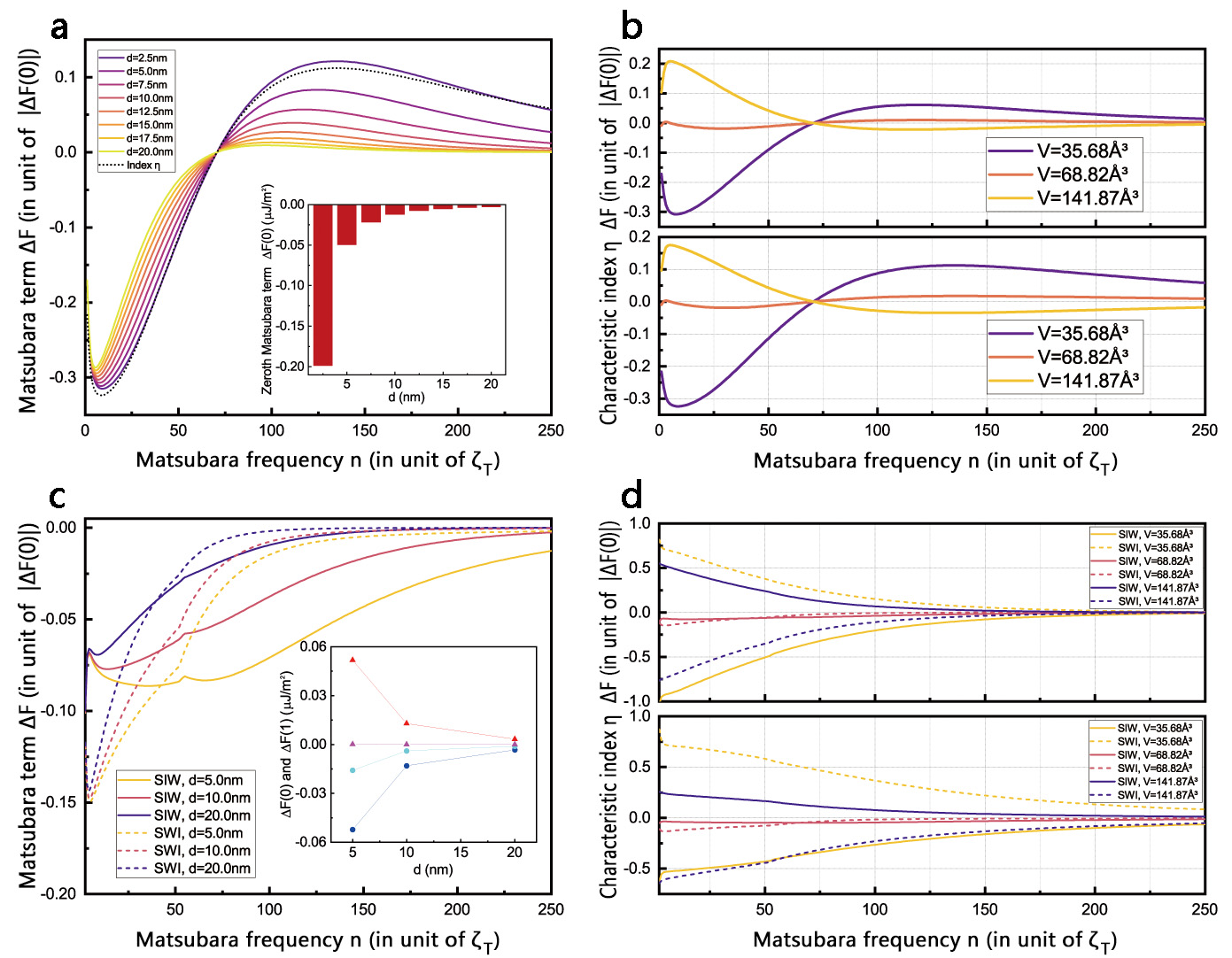}
  \caption{\label{fig.MC}The contribution from Matsubara terms $\Delta F(n)$ and the characteristic indices $\eta$ for silica-ice-water (SIW) in Fig.~\ref{fig.A12} (a and b) and both silica-ice-water and silica-water-ice (SWI) in Fig.~\ref{fig.Fig_EpESLD}c,d. (\textbf{a}) The nonzero Matsubara terms $\Delta F(n),n\geq1$ defined in Eq.~\eqref{eqD.1} normalized by $|\Delta F(0)|$ when the average cell volume of the silica is $V=35.68$\,{\AA}$^3$. (\textbf{b}) A comparison between $\Delta F(n)$ and the characteristic index $\eta$ defined in Eq.~\eqref{eqD.2} with $d=7.0\,\rm nm$ fixed. (\textbf{c}) The Matsubara terms $\Delta F(n)$ defined in Eq.~\eqref{eqD.1} normalized by $|\Delta F(0)|$ when the average cell volume of the silica is $V=68.82$\,{\AA}$^3$. The zeroth (silica-ice-water, blue circle; silica-water-ice, red triangle) and first (silica-ice-water, cyan circle; silica-water-ice, magenta triangle) order are given in the inset. (\textbf{d}) A comparison between $\Delta F(n)$ and the characteristic index $\eta$ defined in Eq.~\eqref{eqD.2} with $d=10.0\rm nm$ fixed.}
\end{figure*}

 Rewrite the Casimir-Lifshitz free energy in Eq.~\eqref{eqT.1} as the sum of the contribution from each Matsubara frequency
\begin{eqnarray}
\label{eqD.1}
F=\sum_{n=0}^{\infty}\Delta F(n),
\end{eqnarray}
then with this expression we look into the details of Casimir-Lifshitz free energy as Fig.~\ref{fig.MC} shows. As demonstrated in Fig.~\ref{fig.MC}a, for small separations, $\Delta F$ can be large enough to make considerable contributions even in very high orders of Matsubara terms. For instance, when $d=2.5\,\rm nm$, $\Delta F$ can be up to $10\%$ of $|\Delta F(0)|$ even
for $n=125$, and deceases slowly, which renders the total Casimir-Lifshitz free energy large and positive. When the thickness is large, such as $d=20.0\,\rm nm$ in the same figure, for high orders $\Delta F$ remains close to zero. The minimum of Casimir-Lifshitz free energy in Fig.~\ref{fig.A12} is accomplished since the negative Matsubara terms are also suppressed by large separations. So in this case, the properties of dielectric functions in the mid-infrared to ultraviolet region is the most important. To understand the connection between the dielectric properties and the behaviors in Fig.~\ref{fig.MC}a, we introduce the following characteristic index
\begin{eqnarray}
\label{eqD.2}
\eta=A\frac{(\varepsilon_3-\varepsilon_2)(\varepsilon_2-\varepsilon_1)}{\varepsilon_2}.
\end{eqnarray}
in which an arbitrary constant factor $A$ for each system, independent of the properties of media, is added. Then according to Fig.~\ref{fig.MC}a and especially Fig.~\ref{fig.MC}b, the correspondence between $\Delta F$ and $\eta$ is straightforward, which facilitates the prediction for the stability properties due to Casimir-Lifshitz interaction. Both Fig.~\ref{fig.MC}a and Fig.~\ref{fig.MC}b justify the outstanding contribution from the zeroth Matsubara term, and both nonretarded and retarded interactions are important. Typically the relation $\sqrt{\varepsilon_2(i\zeta)}\zeta d\sim1$ could be useful when estimating the size of intermediate medium if any. The correspondence between $\Delta F$ and $\eta$ is also evidently illustrated by Fig.~\ref{fig.MC}d. A particularly novel situation, introduced by the latest data for ice and water,\cite{luengo2021lifshitz,luengo2021WaterIce} is the $V=68.82$\,{\AA}$^3$ case depicted in Fig.~\ref{fig.Fig_EpESLD}. Fig.~\ref{fig.MC}c shows that either for silca-ice-water or silica-water-ice, $\Delta F$ in the $V=68.82$\,{\AA}$^3$ case is always negative, which is due to negative (but close to zero) characteristic indices for either case.

\par
\begin{figure}[htp]
  \centering
  \includegraphics[scale=0.32]{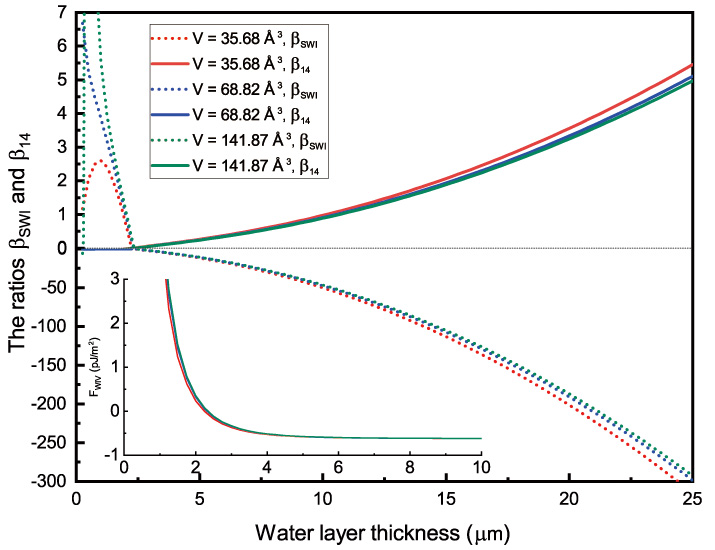}
  \caption{\label{fig.B15}The contributions from different Casimir-Lifshitz interactions to the minimum free energies with different given water layer thicknesses. The systems investigated in Fig.~\ref{fig.FigLD4L} are studied in detail. The contributions from $F_{\rm SWI}$ and $F_{14}$ are shown in terms of $\beta_{\rm SWI}$ and $\beta_{14}$ defined in Eq.~\eqref{eqD.3}. In the inset, the corresponding free energy of water-ice-vapor configuration is given.}
\end{figure}

 The four-layer scenarios are much more complex. But we see a clear pattern in Fig.~\ref{fig.FigES4L} and Fig.~\ref{fig.FigLD4L}, and the competition amongst the three interaction terms should be responsible for it. Before we discuss this issue, it is helpful to see how the four-layer system is changed to a three-layer case. To be more specific, we focus on the systems investigated in Fig.~\ref{fig.FigLD4L}, which leads us to results in Fig.~\ref{fig.B15}. There, for each given thickness of the water layer, we consider the minimum Casimir-Lifshitz free energy reached by varying the thickness of ice layer. Denote the contributions to the free energy from silica-water-ice, water-ice-vapor and silica-vapor as $F_{\rm SWI}$, $F_{\rm WIV}$ and $F_{14}$ respectively, corresponding to 1-2-3, 2-3-4 DLP systems and $F_{14}$ described in Eq.~\eqref{eqT.4}, then we use the ratios defined as
\begin{eqnarray}
\label{eqD.3}
\beta_{\rm SWI}=\frac{F_{\rm WIV}}{F_{\rm SWI}},\ \beta_{14}=\frac{F_{\rm WIV}}{F_{14}},
\end{eqnarray}
which are demonstrated in Fig.~\ref{fig.B15}. As one increases the given thickness of water layer, the contributions from $F_{\rm SWI}$ and $F_{14}$ are both suppressed. It is absolutely natural, since with an enlarged separation between silica and ice, and hence silica and vapor, the interaction between these two materials will surely decay. Then only the water-ice-vapor interaction survives, which is illustrated as the consistency between the inset of Fig.~\ref{fig.B15} and Fig.~\ref{fig.Fig_EpESLD}b. More importantly, Fig.~\ref{fig.B15} shows that when the thickness of water layer is not so large, the Casimir-Lifshitz interaction between silica and vapor dominates, which is clearly shown in Fig.~\ref{fig.B15} with relatively small water layer thickness. Therefore, the interactions mediated by more than one medium can be much more significant than the classical DLP contributions, even with larger separations. This is further confirmed by observing that $F_{\rm SWI}$ decays much faster than $F_{14}$ as a larger water layer thickness is given.
\par
\begin{figure}[htp]
  \centering
  \includegraphics[scale=0.32]{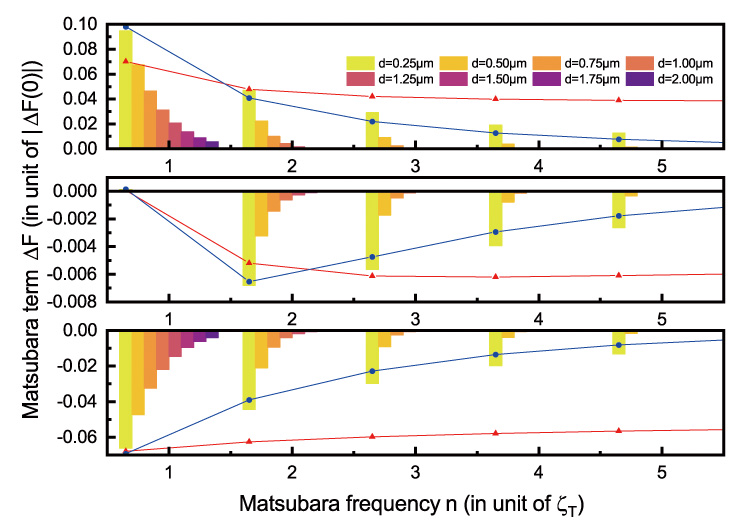}
  \caption{\label{fig.B16}The Matsubara term as a function of the sum $d$ of the water $d_2$ and ice $d_3$ thickness with $d=2d_2$. The systems investigated in Fig.~\ref{fig.FigLD4L} are studied in detail. Red triangles and blue circles are, respectively, plotted with overall scale adjusted for characteristic indices $\eta$ and $\widetilde{\eta}$ defined in Eq.~\eqref{eqD.4} and Eq.~\eqref{eqD.5}, with $d=2d_2=0.25\rm \mu m$.}
\end{figure}

 For the Casimir-Lifshitz interaction mediated by two media, one might expect its free energy in this case here, namely $F_{14}$, could be characterized by the index as in Eq.~\eqref{eqD.2}, that is, from Eq.~(\ref{eqT.4}),
\begin{eqnarray}
\label{eqD.4}
\eta\propto \widetilde{r}_{43}(1-\widetilde{r}_{32}^2)\widetilde{r}_{21},\
\widetilde{r}_{ij}=\frac{\varepsilon_j-\varepsilon_i}{\varepsilon_j+\varepsilon_i}.
\end{eqnarray}
Equation~\eqref{eqD.4} includes the direct interaction between silica-water (or 1-2) and ice-vapor (or 3-4) interfaces and the leading contribution passing through the water-ice (or 2-3) interface once. However, although one can check that contributions from interactions mediated by a third interface is quite small compared to the direct interaction, namely the $O(r_{43}^sr_{21}^s)$ term in Eq.~\eqref{eqT.4} or $\widetilde{r}_{43}\widetilde{r}_{21}$ in the characterized form, Fig.~\ref{fig.B16} shows that the characteristic index defined in Eq.~\eqref{eqD.4} is not enough in this case. As the Matsubara frequency increases, the Matsubara term and its corresponding characteristic index $\eta$ (red triangle) behave distinctly different. This actually illustrates the fact that properties of materials can strongly influence the effective range of Casimir-Lifshitz interaction. Unlike in Fig.~\ref{fig.MC}, the interaction range in the four-layer case here is of micron size order, almost one thousand times larger than that in Fig.~\ref{fig.MC}. It is thus reasonable to include the contribution from the retardation effect, which leads us to a modified characteristic index, defined as follows
\begin{eqnarray}
\label{eqD.5}
\widetilde{\eta}\propto\widetilde{r}_{43}(1-\widetilde{r}_{32}^2)\widetilde{r}_{21}
e^{-2(\sqrt{\varepsilon_2}d_2+\sqrt{\varepsilon_3}d_3)\zeta}.
\end{eqnarray}
In Fig.~\ref{fig.B16}, $\widetilde{\eta}$ (blue circle) proves itself a good characteristic index for the Casimir-Lifshitz free energy mediated by more than one layer at least for $F_{14}$ in Eq.~\eqref{eqT.3}. So both properties of materials and the retardation of field (in addition to the static interaction) exert significant influences. The minimum of the total Casimir-Lifshitz free energy is obtained due to a subtle balance amongst those various factors above, which show themselves in each contributing free energies no matter whether they arise from three- or four-layer cases. The Casimir-Lifshitz interaction is many-body in nature after all.
\par
\begin{figure}[htp]
  \centering
  \includegraphics[scale=0.32]{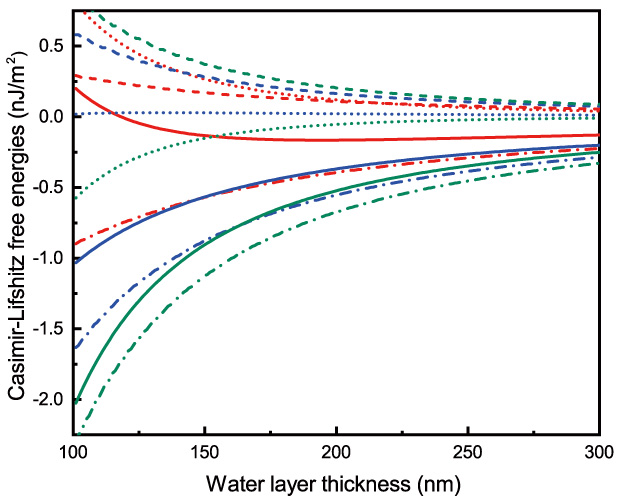}
  \caption{\label{fig.B15details}The systems investigated in Fig.~\ref{fig.FigLD4L} are studied in detail. For different silica materials with the average cell volume $V=35.68${\AA}$^3$ (red), $V=68.82${\AA}$^3$ (blue) and $V=141.87${\AA}$^3$ (green), the Casimir-Lifshitz free energies from silica-water-ice $F_{\rm SWI}$ (dot), water-ice-vapor $F_{\rm WIV}$ (dash), four-layer interaction $F_{14}$ (dash dot), and the total free energy $F=F_{\rm SWI}+F_{\rm WIV}+F_{14}$ (solid) are plotted, when the minimum of $F$ is reached for each given water layer thickness.}
\end{figure}

 Though complicated, we could also try to understand some of the details of this subtle balance from both Fig.~\ref{fig.FigES4L} and Fig.~\ref{fig.FigLD4L}, in which a clear pattern presents, though they look complex at first glance. Fig.~\ref{fig.FigES4L}a obviously gains its two trenches corresponding to the minima of Casimir-Lifshitz free energies for silica-ice-water and ice-water-vapor evaluated with the data in previous work.\cite{MBPhysRevB02017} The similarities between Fig.~\ref{fig.FigES4L}c,e and Fig.~\ref{fig.FigLD4L}a,c are also apparent. For simplicity, we focus on the systems depicted by Fig.~\ref{fig.FigLD4L} again. According to Fig.~\ref{fig.B15details}, for the silica material with the average cell volume $V=35.68${\AA}$^3$, both silica-water-ice $F_{\rm SWI}$ and water-ice-vapor $F_{\rm WIV}$ contribute positively when the thickness of water layer is given in the range $100\rm\ nm<d_2<300\rm\ nm$. On the contrary, $F_{14}$ is always negative, and together $F_{\rm SWI}$ and $F_{\rm WIV}$ produces a global minimum of total Casimir-Lifshitz free energy at about $d_2\approx0.2\rm\ \mu m$, just as shown in Fig.~\ref{fig.FigLD4L}a. Larger average cell volume decreases the permittivity of silica material, which not only renders $F_{14}$ more negative, but also eliminates the ability of $F_{\rm SWI}$ to compensate $F_{14}$. So the behaviors shown in Fig.~\ref{fig.FigLD4L}c,e are mainly caused by $F_{14}$ (also see the $V=68.82${\AA}$^3$ and $V=141.87${\AA}$^3$ cases in Fig.~\ref{fig.B15details}). Based on these analyses, it can be expected that with the characteristic indices for DLP and four-layer interactions defined above, the possible influences of Casimir-Lifshitz interaction on the stability of a configuration under study can be predicted qualitatively.

\section{Conclusions}
\label{C}
\par In this paper, we redo and extend the previous work on the premelting and formation of ice, either on a bulk of silica rock or on the ice-cold water, caused by the Casimir-Lifshitz interaction. A mistake in our previous work\cite{MBPhysRevB02017} has been spotted and the corrected results are given. It can be safely claimed that the detailed properties of materials (in the cases considered in this paper, their permittivities) influence these premelting and formation phenomena. Perhaps the most important observation from the recent modelling of ice and water could be that their dielectric functions at the triple point of water only have crossings between the $n=0$ and $n=1$ Matsubara frequencies.\cite{JohannesWater2019,luengo2021lifshitz,luengo2021WaterIce}
This fundamentally alters the predicted behaviour compared to when the dielectric functions from Elbaum and Schick are used.\cite{elbaum1991app} For the original ice-water-vapor structure, the most recent dielectric functions for ice and water lead to the formation of a thin ice layer on the water, which is in contradiction with the premelting prediction from Elbaum and Schick.\cite{elbaum1991app} Our results confirm those in Ref.~\onlinecite{JohannesWater2019}, that is, micron-sized ice layer forms on the surface of cold water. We suggest that careful optical measurements of water and ice at different temperatures and pressures should be highly interesting. Such measurements have the potential to improve our understanding of the role for dispersion forces in ice formation and melting.

\par As can be expected, the three-layer silica-ice-water and four-layer silica-ice-water-vapor systems, behave differently from those studied in previous work using the parameters from Elbaum and Schick.\cite{elbaum1991app,Elbaum2,Wilen,Thiyam2016,Bostr2016,MBPhysRevB02017,Prachi2019role,esteso2020premelting} The ice formation on the silica is not seen, but a bulk of ice or water is directly in contact with the silica surface, depending on the permittivity of the particular silica materials or the initial condition of ice or water for some cases, such as the $V=68.82$\,{\AA}$^3$ case here.

\par In four-layer scenarios, we see both a single global stable state of water and ice layers (Fig.~\ref{fig.FigES4L}c, Fig.~\ref{fig.FigLD4L}a), the coexistence of a meta-stable and a stable state (Fig.~\ref{fig.FigES4L}a), and the relatively stable region like a waterfall (Fig.~\ref{fig.FigES4L}e, Fig.~\ref{fig.FigLD4L}c,e). Those complicated situations occur due to correlations of different types of Casimir-Lifshitz interactions. Two indices are introduced to characterize those interactions here for the four-layer case. With the characteristic indices, it will be easier to estimate influences of the Casimir-Lifshitz interaction on the stability of the system to be investigated. On the theoretical side, they and their generalizations in more complex situations, for instance multi-slab cases in general, may help to throw light on the mechanics of the Casimir-Lifshitz interaction mediated by inhomogeneous materials, which is now still largely veiled.

\par In this work we have analyzed the induced inhomogeneous free energies
in layered media generated by Casimir-Lifshitz interactions.  This
analysis assumes that surface charge and surface adsorption of free
ions (salt or dissociated water) is not significant, corresponding to
a system at the isoelectric point (IEP) where any such surface charge
is neutralized. Both ice \cite{KallayCopChibowskiHolysz2003} and
quartz (silica \cite{HartleyLarsonScales1997}) have a low IEP around
pH 3--3.5. At other pH conditions away from the IEP, including pH 7,
charge-dependent effects can be
important\,\cite{ThiyamFiedlerBuhmannPerssonBrevikBostromParsons2018},
with ion adsorption layers generally creating a repulsive force that
would enhance premelting.  These charge-induced interactions deserve
further attention and must in general be important for both ice
formation and premelting.

\begin{acknowledgments}
The work of KAM was supported in part by a grant from the US National Science Foundation, No.~2008417. We acknowledge support from the Research Council of Norway (Projects No.~221469 and No.~250346).
We also acknowledge access to high-performance computing resources via SNIC and NOTUR.
Last, we would especially like to thank Juan Luengo M\'arquez and Dr. Luis G. MacDowell who shared their independently prepared files with dielectric function data\citep{luengo2021lifshitz,luengo2021WaterIce} for ice and water.
\end{acknowledgments}

\appendix

\section{Derivation of free energy for four-slab system}
\label{AppFL}
We consider a 4-layer system, each layer being homogeneous:
\be
\varepsilon,\mu(z)=\left\{\begin{array}{cc}
z<a:&\epsilon_1,\mu_1,\\
a<z<b:& \epsilon_2,\mu_2,\\
b<z<c:& \epsilon_3,\mu_3,\\
c<z:& \epsilon_4,\mu_4.
\end{array}\right.
\ee
The free energy is given in general by
\be
F=\frac{T}2
\sum_{n=-\infty}^\infty \int\frac{d^2k}{(2\pi)^2}\ln\Delta^E\Delta^H,
\ee
the sum being over Matsubara frequencies $\zeta_m$.
We will use the inhomogeneous medium description given in Ref.~\onlinecite{LiMiltonGuoKennedyFulling2019}.
We regard the regions 2 and 3 as a single region, called ``in'' where the
medium is inhomogeneous.  To obtain an unambiguously finite result, we subtract
a reference energy corresponding to removing the boundary $a$, that is, letting
medium 2 to extend to $-\infty$.  Then, we add back in the energy corresponding
to that reference energy.  So
\be
F=F_{\rm sub}+F_{\rm ref}.
\ee
  Here we consider the TE contribution only, the
TM contribution being obtained by the obvious substitutions.

Consider first the reference situation, which is just the familiar DLP
configuration.  There, according to Ref.~\onlinecite{LiMiltonGuoKennedyFulling2019}
\be
\Delta^E_{\rm ref}=1-\frac{[e_{2,-},e_{3,-}]_\mu(b)[e_{3,+},e_{4,+}]_\mu(c)}
{[e_{2,-},e_{3,+}]_\mu(b)[e_{3,-},e_{4,+}]_\mu(c)},
\ee
where $e_i$ satisfy
\be
\left(\partial_z\frac1{\mu_i}\partial_z-\frac{k^2}{\mu_i}-\varepsilon_i\zeta^2\right)
e_i=0,\label{dee}
\ee
in which the medium $i$ has been analytically extended to the whole space. The generalized Wronskians are defined by (prime means derivative with respect to the argument)
\be
[e_i,e_j]=\frac1{\mu_i} e_i'e_j-\frac1{\mu_j}e_j'e_i,
\ee
evaluated at the same point.

In the DLP configuration, we may define
\be
e_{i,\pm}=e^{\mp\kappa_i z},\quad \kappa_i=\sqrt{k^2+\epsilon_i\mu_i\zeta^2}.
\ee
Then it is immediate to find
\be
\Delta^E_{\rm ref}=1-r^E_{23}r^E_{43}e^{-2\kappa_3(c-b)},\label{deltaref}
\ee
in terms of the reflection coefficients
\be
r^E_{ij}=\frac{\hat\kappa_j-\hat\kappa_i}{\hat\kappa_j+\hat\kappa_i},\quad
\hat\kappa_i=\frac{\kappa_i}{\mu_i}.
\ee
This directly gives the Lifshitz energy in Eq.~\eqref{eqT.1}.

Now for the subtracted four-slab configuration, we need to compute
\be
\Delta^E_{\rm sub}=1
-\frac{[e_{1,-},e_{\rm{in},-}]_\mu(a)[e_{\rm{in},+},e_{4,+}]_\mu(c)}
{[e_{1,-},e_{\rm{in},+}]_\mu(a)[e_{\rm{in},-},e_{4,+}]_\mu(c)},
\ee
where the effort is only in finding the solution in the $2+3$ region.
We can take $e_{\rm{in}\mp}(z)$ to be $e^{\pm\kappa_2 z}$ for $a<z<b$; then by
requiring, from the differential equation Eq.~\eqref{dee}, continuity of the
solution, and of $\frac1{\mu}$ times the derivative of the solution, we find
\begin{subequations}
\begin{eqnarray}
e_{\rm{in},\mp}
&=&
e^{\pm\kappa_2 z},\ a<z<b
\end{eqnarray}
\begin{eqnarray}
e_{\rm{in},\mp}(z)
&=&
\bigg[(\hat\kappa_3-\hat\kappa_2)
e^{\mp\kappa_3(z-b)}
+(\hat\kappa_3+
\hat\kappa_2)e^{\pm\kappa_3(z-b)}\bigg]
\nonumber\\
& &
\times\frac{1}{2\hat\kappa_3}e^{\pm\kappa_2b},\
b<z<c.
\end{eqnarray}
\end{subequations}
Then $\Delta^E_{\rm sub}$ is readily calculated to be
\begin{eqnarray}
\Delta^E_{\rm sub}=1+r_{12}^Ee^{-2\kappa_2(b-a)}\frac{r_{23}^E e^{2\kappa_3(c-b)}
+r_{34}^E}{e^{2\kappa_3(c-b)}+r_{34}^E r_{23}^E} .\label{deltasub}
\end{eqnarray}
When this is multiplied by $\Delta^E_{\rm ref}$ in Eq.~(\ref{deltaref}), the
denominator in Eq.~(\ref{deltasub}) is cancelled, and we are left with
\begin{eqnarray}
\Delta^E
&=&
1+r^E_{43}r^E_{32}e^{-2\kappa_3(c-b)}+r_{32}^Er_{21}^E e^{-2\kappa_2(
b-a)}
\nonumber\\
& &
+r_{21}^E r_{43}^Ee^{-2\kappa_2(b-a)}e^{-2\kappa_3(c-b)},
\end{eqnarray}
which is precisely the result stated in Eq.~(4) after we replace $b-a \to d_2 $, $c-b \to d_3$. The decomposition given in Section II of $F=F_{14}+F_{13}+F_{24}$, is now
immediate, where $F_{14}$ has a rather immediate interpretation in terms of multiple
scattering, given that the transmission coefficient between region 2 and 3 satisfies
$t_{23}^2=1-r_{23}^2$.

\bibliography{ref}

\end{document}